\documentclass{ws-procs11x85}
\usepackage{ws-procs-thm} 
\usepackage{booktabs}
\usepackage{tabularx}
\usepackage[table]{xcolor}
\usepackage[colorlinks=false]{hyperref}
\newcommand{\colorrows}{\rowcolors{2}{white}{cyan!15}}

\begin{document}

\title{Deciphering the influence of demographic factors on the treatment of pediatric patients in the emergency department}

\author{Helena Coggan\textsuperscript{1,2}, Anne Bischops\textsuperscript{1,2}, Pradip Chaudhari\textsuperscript{3}, Yuval Barak-Corren\textsuperscript{4}, Andrew M. Fine\textsuperscript{2,5}, Ben Y. Reis\textsuperscript{1,2}, Jaya Aysola\textsuperscript{6} and William G. La Cava\textsuperscript{1,2}}

\address{\textsuperscript{1}Computational Health Informatics Program, Boston Children's Hospital, Boston, MA, USA}
\address{\textsuperscript{2}Harvard Medical School, Boston, MA, USA}
\address{\textsuperscript{3}Division of Emergency and Transport Medicine, Children's Hospital Los Angeles and Department of Pediatrics, Keck School of Medicine of the University of Southern California, Los Angeles, CA, USA}
\address{\textsuperscript{4}Department of Pediatric Cardiology, Schneider Children's Medical Center, Affiliated to Tel Aviv University Faculty of Medical and Health Sciences, Petach Tikvah, Israel}
\address{\textsuperscript{5}Division of Emergency Medicine, Boston Children's Hospital, Boston, MA, USA}
\address{\textsuperscript{6}Leonard Davis Institute of Health Economics, University of Pennsylvania; Department of Medicine, Perelman School of Medicine, University of Pennsylvania; and Penn Medicine Center for Health Equity Advancement, Philadelphia, Pennsylvania}

\begin{abstract}
Persistent demographic disparities have been identified in the treatment of patients seeking care in the emergency department (ED). These may be driven in part by subconscious biases, which providers themselves may struggle to identify. To better understand the operation of these biases, we performed a retrospective cross-sectional analysis using electronic health records describing 339,400 visits to the ED of a single US pediatric medical center between 2019-2024. Odds ratios were calculated using propensity-score matching. Analyses were adjusted for confounding variables, including chief complaint, insurance type, socio-economic deprivation, and patient comorbidities. We also trained a machine learning [ML] model on this dataset to identify predictors of admission. We found significant demographic disparities in admission (Non-Hispanic Black [NHB] relative to Non-Hispanic White [NHW]: OR 0.77, 95\% CI 0.73-0.81; Hispanic relative to NHW: OR 0.80, 95\% CI 0.76-0.83). We also identified disparities in individual decisions taken during the ED stay. For example, NHB patients were significantly less likely than NHW patients to be assigned an `emergent' triage acuity score of (OR 0.70, 95\% CI 0.67-0.72), but emergent NHB patients were also significantly less likely to be admitted than NHW patients with the same triage acuity (OR 0.86, 95\% CI 0.80-0.93). Demographic disparities were particularly acute wherever patients had normal vital signs, public insurance, moderate socio-economic deprivation, or a home address distant from the hospital. An ML model assigned higher importance to triage score for NHB than NHW patients when predicting admission, reflecting these disparities in assignment. We conclude that many visit characteristics, clinical and otherwise, may influence the operation of subconscious biases and affect ML-driven decision support tools. 
\end{abstract}


\copyrightinfo{\copyright\ 2025 The Authors. Open Access chapter published by World Scientific Publishing Company and distributed under the terms of the Creative Commons Attribution Non-Commercial (CC BY-NC) 4.0 License.}
\section{Background}

Demographic disparities in emergency care and outcomes are an ongoing concern\cite{Nelson2002, Hamberg2008GenderMedicine}. 
Studies of adult patients show that Black patients are assigned lower triage acuity scores (ESI) \cite{Schrader2013} than White patients, and Black and Hispanic patients with non-urgent triage acuity wait longer to be seen \cite{Lu2021}.
Black patients admitted to hospital also experience longer ED stays than White patients \cite{Pines2009}.  
These disparities appear in the treatment of specific conditions \cite{Martin2024,Longcoy2022}, and in outcomes beyond admission, such as lower likelihood of receiving opioids for pain-related visits~\cite{Kang2022}. 
Gender-based differences are also evident: women more often report difficulty obtaining timely ED care\cite{Chen2022}. 

These disparities extend to the pediatric emergency care setting. 
Pediatric patients face racial differences in triage assessments, both overall \cite{Metzger2022} and for specific complaints such as fever \cite{Dennis2021}. 
Black, Hispanic, Asian and Native American pediatric patients are less likely than White patients to receive radiological tests, though gaps narrow for conditions with standardized protocols~\cite{Payne2009}. 
Providers are also more likely to interrupt, and less likely to make eye contact with, patients from different racial and ethnic groups~\cite{Aysola2022}, suggesting subconscious bias influences ED care well before admission decisions. 
Potentially biased upstream decisions include triage acuity scores, test orders, and prioritizing care to one patient before another~\cite{Sangal2023SociodemographicCare}. 
Biases may also affect clinical interpretation of vital signs and non-clinical variables such as insurance type and socio-economic status, which in turn affect the output of clinical decision support tools trained to predict admission~\cite{lacavaFairAdmissionRisk2023a,Barak-Corren2017,Barak-Corren2021} or clinical orders~\cite{hunter-zinckPredictingEmergencyDepartment2019}. 
To reduce sex- and race-based disparities, and to safely deploy predictive algorithms, we need a fuller picture of the mechanisms driving ED decision-making.

In this study we quantified demographic disparities in admission at a large urban pediatric ED in Boston, Massachusetts. 
We considered two main classes of disparity. 
First, we examined \emph{upstream disparities}: differences in clinical decisions made before decisions to admit or discharge (e.g.  the likelihood of receiving a medication or triage acuity score). 
Second, we examined \emph{downstream disparities}: differences in outcomes like admission rates between patients with the same medication or score. 
Finally, we assessed \emph{feature importance disparities}: differences in the predictive importance an ML model assigns to visit characteristics across demographic groups.

Our study provides a more detailed view of disparities within a hospital than is possible using the widely-studied National Hospital Ambulatory Medical Care Survey (NHAMCS)~\cite{NationalHospitalAmbulatory2017}, which lacks some key variables. 
These results can inform clinical education and help developers design fairer clinical support tools.

\section{Study Sample and Available Information}

We performed a retrospective study of 352,324 visits to a large urban pediatric ED between January 2019 and December 2024. We excluded 3,452 visits (0.98\%) which lacked demographic information; 5,606 (1.6\%) which did not result in admission or discharge; and 3,149 (0.9\%) which did not have a linked chief complaint. 
A small number of visits ($<$0.5\%) were further excluded because of recording errors (obviously implausible lengths of stay in months or years, departure or admission request preceding arrival, no recorded sex, or no consistent patient identifier). 
Where a final disposition was not recorded, we inferred admission from the presence of an admission request, and inferred discharge if vitals were recorded but no admission request was made. 
Our final dataset contained 339,400 visits (96\% of the original dataset) from 179,944 unique patients. The overall composition of these visits is recorded in Table \ref{tab:demographics}.

Our primary exposures of interest were sex (male or female) and race (Asian, Hispanic, Non-Hispanic Black, Hispanic White, Non-Hispanic White, Other,  and unknown). Available data included age (categorised into 0-3 months, 3-6 months, 6-12 months, 12-18 months, 18-36 months, 3-5 years, 5-10 years, 10-15 years, and 15+ years \cite{Barak-Corren2017}), primary language (condensed to English, Spanish or Other), and type of insurance (public or private). We also considered the `triage acuity score' each patient was assigned shortly after arrival (also known as the Emergency Severity Index or ESI score). A score of 1 indicates that a patient requires immediate and lifesaving intervention, whereas a score of 5 indicates low severity.

From the patient's zip code, we calculated the distance travelled to the ED for the visit (assuming that a patient left from their home address), whether a patient came from out of state, and a socio-economic deprivation index (SDI), using the scores developed by the Robert Graham Center for 2019, the most recent year available \cite{GrahamCenter} (see Table \ref{tab:quartiles}). 

We also included a patient's history within the thirty days preceding the visit, separated into number of previous admissions and number of previous visits without admission. Other visit-specific variables included mode of transport to the ED; season, time of day, year of arrival, and whether the visit occurred on a weekend; and the number of other patients in the ED at the time a patient arrived (referred to throughout this paper as `crowdedness'), weighted by severity (so that a patient with an ESI score of 1 was weighted 5 times more than a patient with an ESI score of 5). Chief complaint was recorded as a free text field, which complicated the consistent identification of ailments. To deal with this, we first manually expanded common abbreviations (`s.o.b.' to `shortness of breath'), and then shortened each word in a patient's complaint to its `stem' (e.g. `diabetes' and `diabetic' to `diabet') using the R package SnowballC \cite{SnowballC}. We then categorised complaints using the Pediatric Emergency Reason for Visit clustering system \cite{Gorelick2005}.

We also had information on a patient's vitals across their ED stay: heart rate (HR), respiration rate (RR), oxygen saturation, blood pressure, and reported pain (on a scale of 0-10, with 0 corresponding to `no pain'). We recorded the minimum, maximum and mean reading of each vital across a visit as separate variables. HR, RR and weight variables were converted to a score corresponding to the number of standard deviations from the mean of each age group \cite{Barak-Corren2017}. Blood pressure scores were calculated according to the PALS (Pediatric Advanced Life Support) criterion, which incorporates systolic blood pressure (SBP) and age to determine whether a pediatric patient is hypotensive.

We controlled for a patient's medical history, and for the diagnoses resulting from the ED visit, by defining two variables for each category of comorbidities in the pediatric comorbidity index (PCI) \cite{Sun2021}.
For each category (e.g. developmental disorders, chromosomal abnormalities, ), we recorded the presence of a previous condition if a relevant diagnosis had been noted more than one day but less than one year before the visit, and a `current condition' if a relevant diagnosis was noted within a day of the visit. (This does not necessarily mean that the diagnosis was \textit{initially made} during the visit, but it does imply that the diagnosis was relevant to that visit).

\begin{table}
    \tbl{Demographic characteristics of the study population in the BCH dataset. Numbers correspond to visits. ESI: Emergency Severity Index. EMS: Emergency medical services (ambulance, helicopter, etc).}
    {
\footnotesize
\colorrows
    \begin{tabularx}{0.97\textwidth}{l l r r}\toprule
    \rowcolor{white}
          Characteristic&&  Number (\%) of visits& Admission rate (\%)\\
          \midrule
          \cellcolor{white}Sex&Female&  161725 (47.7)& 17.4\\
          \cellcolor{white}&Male&  177675 (52.3)& 18.4\\
          \cellcolor{cyan!15} Race
            \cellcolor{cyan!15}&Asian&  13260 (3.91)& 18.3\\
            \cellcolor{cyan!15}&Hispanic&  88106 (26)& 12.2\\
            \cellcolor{cyan!15}&Hispanic White&  728 (0.214)& 17.9\\
            \cellcolor{cyan!15}&Non-Hispanic Black&  50853 (15)& 12.5\\
            \cellcolor{cyan!15} &Non-Hispanic White&  116339 (34.3)& 25.5\\
            \cellcolor{cyan!15} &Other&  33639 (9.91)& 16.4\\
          \cellcolor{cyan!15}   &Unknown&  36475 (10.7)& 16.5\\
          \cellcolor{white} Age
            &0-3 months &  14178 (4.18)& 30\\
          \cellcolor{white}&3-6 months&  9332 (2.75)& 20.3\\
          \cellcolor{white}&6-12 months&  20089 (5.92)& 15\\
          \cellcolor{white}&12-18 months&  18300 (5.39)& 14.4\\
          \cellcolor{white}&18-36 months&  41084 (12.1)& 14\\
          \cellcolor{white}&3-5 years&  40078 (11.8)& 14.2\\
          \cellcolor{white}&5-10 years&  69520 (20.5)& 15.8\\
          \cellcolor{white}&10-15 years&  58468 (17.2)& 19.7\\
          \cellcolor{white}&15 years and older&  68351 (20.1)& 22.1\\
          \cellcolor{cyan!15}
          Mode of arrival &EMS&  27295 (8.04)& 31.7\\
          \cellcolor{cyan!15}&Transfer&  14429 (4.25)& 55.1\\
          \cellcolor{cyan!15}&Walk in&  296449 (87.3)& 14.8\\
          \cellcolor{cyan!15}&Unknown&  1227 (0.362)& 24.3\\
          \cellcolor{white}Primary Language&Arabic&  4734 (1.39)& 22.7\\
          \cellcolor{white}&Cape Verdean&  1819 (0.536)& 12.2\\
          \cellcolor{white}&Chinese Mandarin&  1357 (0.4)& 15.1\\
          \cellcolor{white}&English&  270314 (79.6)& 19\\
          \cellcolor{white}&Haitian Creole&  3071 (0.905)& 13.4\\
          \cellcolor{white}&Other&  13104 (3.86)& 15.5\\
          \cellcolor{white}&Portuguese&  6648 (1.96)& 17.6\\
          \cellcolor{white}&Spanish&  38353 (11.3)& 11.2\\
          \cellcolor{cyan!15}Triage acuity (ESI) score&1&  1535 (0.452)& 71.9\\
          \cellcolor{cyan!15}&2&  83163 (24.5)& 41.2\\
          \cellcolor{cyan!15}&3&  159571 (47)& 15.4\\
          \cellcolor{cyan!15}&4&  85969 (25.3)& 0.943\\ 
          \cellcolor{cyan!15} & 5& 8811 (2.6)&0.375\\
          \cellcolor{cyan!15}    & Unknown& 351 (0.103)&20.8\\
        \cellcolor{white}Insurance& Public& 167458 (49.3)&21.1\\
         & Private& 171942 (50.7)&14.8\\ 
         \bottomrule
    \end{tabularx}}
    \label{tab:demographics}
\end{table}

\begin{table}
    \tbl{Distributions of the features of visits to the BCH ED. Distance travelled to the hospital is calculated as the distance between the patient's zip code and the hospital's (02115). A higher social deprivation score indicates a poorer zip code.}
    {\begin{tabular}{@{}lccc@{}}\toprule
          Characteristic&  Lower quartile&  Median& Upper quartile\\\colrule
          Length of stay (minutes)&  158&  244& 372\\
         Distance travelled to hospital (miles)&  3.78&  7.44& 21.69\\
 Social deprivation index& 18& 57&92\\
 \bottomrule
    \end{tabular}}
    \label{tab:quartiles}
\end{table}

\section{Statistical Analyses}

As a majority of visits were from male patients and a plurality from Non-Hispanic White (NHW) patients (see Table \ref{tab:demographics}), male sex and NHW race were taken as reference groups. Our primary outcome of interest was admission to hospital. Secondary outcomes included assignment of ESI score, dispensing of medications, ordering of radiological tests, and length of stay. 

We calculated odds ratios using propensity-score matching. Propensity scores were calculated using the R package glmnet \cite{glmnet1, glmnet2}. If a sociodemographic variable was unknown (e.g. SDI), we recorded this in a separate binary variable (e.g. SDI\_unknown) and set the value of the original variable for that visit to the cohort mean (on the basis that nothing could be known about the patient's socioeconomic deprivation beyond what is known about the population served by the ED as a whole).

For each demographic characteristic (e.g. Non-Hispanic Black [NHB] race), we restricted the cohort to this demographic and the reference group (e.g. only visits from NHB and NHW patients) and fitted a logistic regression model to `predict' NHB race using all relevant variables. Due to the large number of variables available (around 300), we fit propensity scores using LASSO regression. We used five-fold cross-validation and chose the penalty parameter to produce the most regularised model (i.e. to select the fewest variables) whilst remaining within one standard deviation of the minimum possible error. 

When the outcome of interest was triage acuity (ESI score), we excluded all variables from the propensity score which we assumed were not known at triage. These included triage acuity score itself, pediatric comorbidity score, and all `post-triage' vitals. (`Triage vitals' were assumed to be the first set of vitals recorded.) It was not clear when a patient was asked for their insurance type or ZIP code, and so it is possible that some other variables are also unknown at triage (specifically insurance type, miles travelled, SDI, and state of origin). However, we assumed that the triage nurses knew or could guess these variables, and so included them in order to `err on the side of caution' when attempting to measure the independent effect of race and sex on triage acuity (separately from insurance or socio-economic class). 

We matched visits 1:1 with greedy nearest-neighbour matching on the propensity score, using the R package MatchIt \cite{MatchIt}. We set a caliper threshold of 10\% of the standard deviation of the propensity score, and recalculated at 20\% to check that our results were robust to caliper thresholds. Odds ratios were calculated using McNemar's exact test. Multiple hypothesis correction was conducted using the Bonferroni method. For each set of outcomes (overall admission, upstream decisions, and admission following upstream decisions)  we performed Bonferroni correction across all interventions simultaneously. 

We also sought to test the effect of downstream disparities on the quality of predictive algorithms trained on these data. To do this, we designed a model using XGBoost \cite{XGBoost} to predict admission on a continuous basis during a patient's stay. Stays were `sampled' at the start and end of their visit, and then every 60 minutes within the visit itself. At each sampled timepoint, all information known about the patient up to this timepoint was collated, including the ordering of any radiological tests, the dispensing of the top 50 most common medications (including the method of its administration, e.g. intravenous [IV] or oral), and the most recent results of the 50 most common laboratory tests (age-normalised so that each result lies between 0 and 1, with the 1st and 99th percentile set as lower and upper bounds for stability). As the timing of triage assessment could not be reliably determined, we assumed that ESI score and demographic characteristics were available at every timepoint during the patient's stay. We took the first three and a half years of visits as the training data and the next year (roughly May 2023-May 2024) as the test data (removing visits after this point due to difficulties in mapping radiological tests). Although our train/test split was chronological, we included the year of admission as features in the test and training data to allow the model to learn from changes in admission rates during the pandemic. We split before extracting timepoint-samples, to ensure information about the same visit did not appear in both. Hyperparameters were tuned on a dataset comprising one randomly chosen sample from each visit in the training dataset, using randomised search on 100 hyperparameter sets and five-fold cross validation. During both the training and testing process, timepoints were weighted in inverse proportion to visit length (such that the weights of all timepoints associated with the same visit summed to 1).

\section{Results}

\subsection{Admission to Hospital }

\begin{figure}[htb!]
\centering
\includegraphics[width=\linewidth]{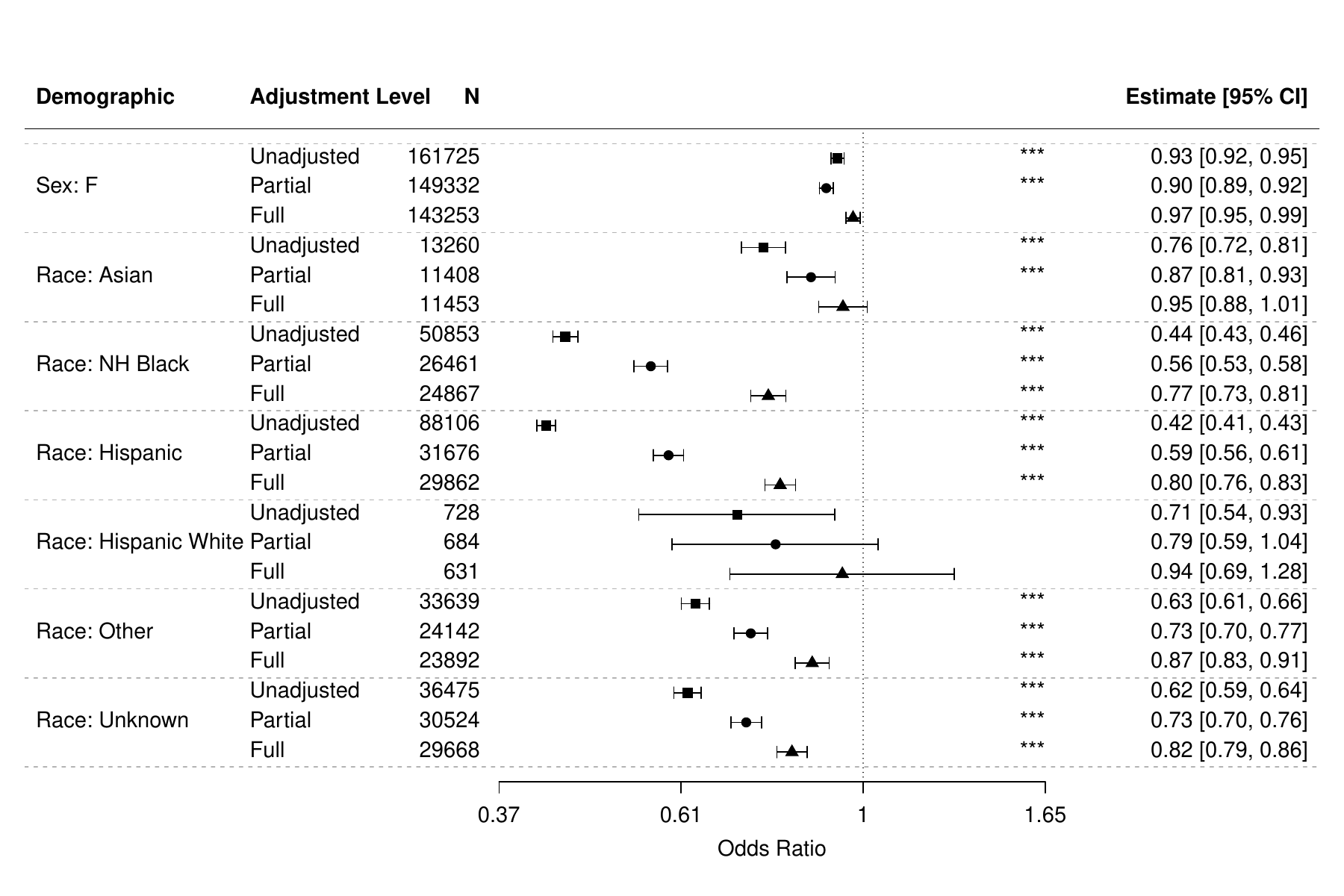}
\caption{Adjusted odds ratios describing the likelihood that different demographic groups of patients will be admitted to hospital. Significance is indicated with asterisks, where p-values have been Bonferroni-corrected across all demographic groups and levels of adjustment simultaneously (*** = p \( < \) 0.001, ** = p \( < \) 0.01, * = p \( < \) 0.05). NH=Non-Hispanic. N=number of visits (i.e. N visits from female patients are compared to N visits from male patients). `Partial' adjustment accounts for all patient-specific factors; full adjustment incorporates visit-specific factors (see text).}
\label{fig:admission-or-plot}
\end{figure}

Visits to the ED during this period had an overall admission rate of 17.9\% and a median LOS of just over 4 hours. We examined the overall impact of race and sex on a patient's odds of being admitted to hospital. We calculated odds ratios at three levels of adjustment. Firstly, we calculated `unadjusted' odds ratios (ORs), where (for example) visits from female patients were matched to visits from male patients at random, without accounting for any other factors. Next, `partially adjusted' ORs were computed using a propensity score which included `patient-specific factors': age group, primary language, history of previous visits, gender identity, socio-economic deprivation, insurance, distance lived from the hospital, state of origin, previous diagnoses, and weight. Partially adjusted odds ratios also accounted for sex when comparing racial groups, and for race when comparing sexes. Finally, we computed `fully adjusted' ORs. In addition to the factors included in the partial adjustment, these accounted visit-specific factors: chief complaint, ESI score, vitals measured across a patient's stay, current diagnoses, mode and time of arrival, and crowdedness.

When performing both partial and full adjustment, covariate balance on the matched cohort was checked using a Love plot. Full balance (standardised mean differences of less than 10\% in all covariates \cite{Austin2011}) was achieved on all cohorts except when comparing visits from Hispanic White (HW) patients to visits from NHW patients, as groups of a few hundred visits are difficult to balance across many covariates.

Unadjusted odds ratios suggested that sex and race had significant effect on hospital admission (see Figure \ref{fig:admission-or-plot}). Before adjustment, visits from all racial groups considered (except HW) were significantly less likely than visits from NHW patients to result in an admission, as were visits from female patients relative to visits from male patients. After full adjustment, significant negative effects could still be observed for all racial groups except HW and Asian. Fully adjusted effects were largest for Hispanic and NHB patients. Racial disparities, and their significance levels, were robust to a doubling of the caliper threshold used to match patients (Hispanic relative to NHW: OR = 0.78, 95\% CI 0.75-0.81; NHB relative to NHW: 0.76[0.73-0.80]). Code used to perform our analyses is available at \url{https://github.com/hcoggan/BCH-ED}.

\subsection{Triage Acuity and Length of Stay} \label{triage-and-ed-los}

We next examined the effect of race on two secondary outcomes: triage acuity and length of stay in the emergency department (LOS). We considered both the odds that a patient would be `assigned' a certain outcome (e.g. an ESI score of 2, or a lowest-quartile LOS) and the odds that a patient \textit{who had been assigned that outcome} would then be admitted. For simplicity, we focused on the three largest demographic groups (other than the reference groups): female patients, Hispanic patients, and NHB patients. All odds ratios were fully adjusted.

\begin{figure}[htb]
\centering
\includegraphics[trim={0cm 0.75cm 0cm 2.2cm},clip, width=0.6\linewidth, height=6cm]{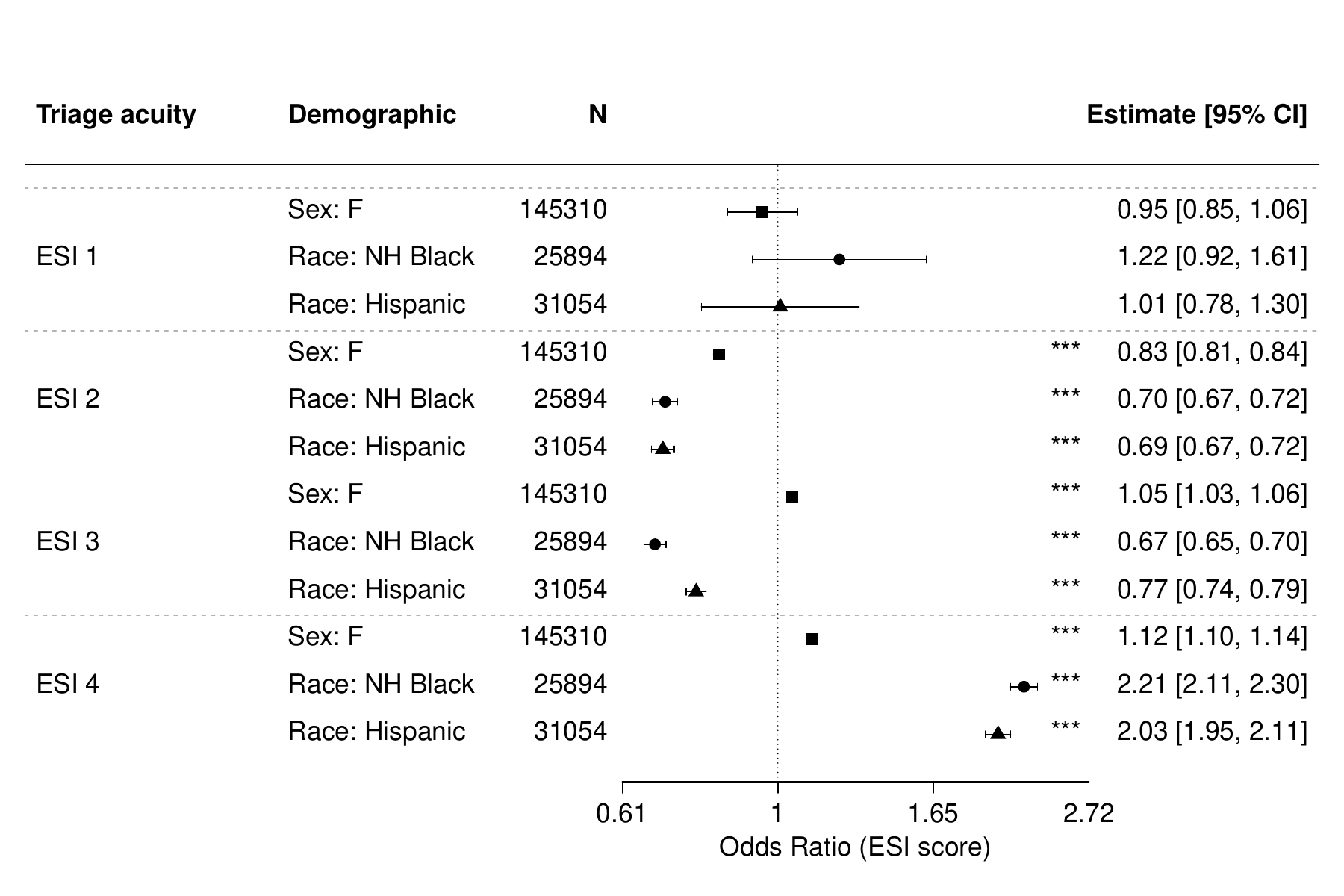}
\includegraphics[trim={12cm 0.75cm 0cm 2.2cm},clip, width=0.39\linewidth, height=6cm]{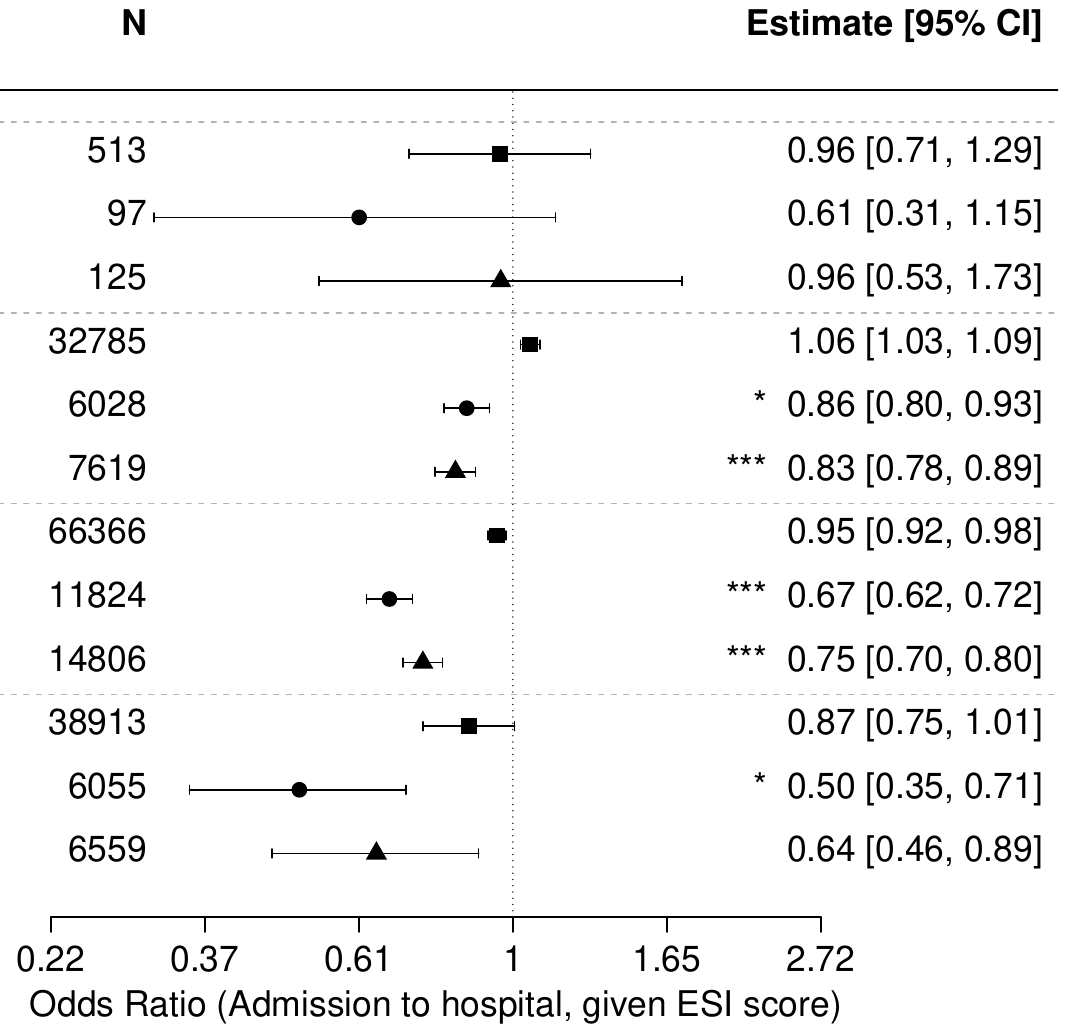}
\includegraphics[trim={0cm 0.5cm 0cm 2.0cm},clip, width=0.6\linewidth, height=6cm]{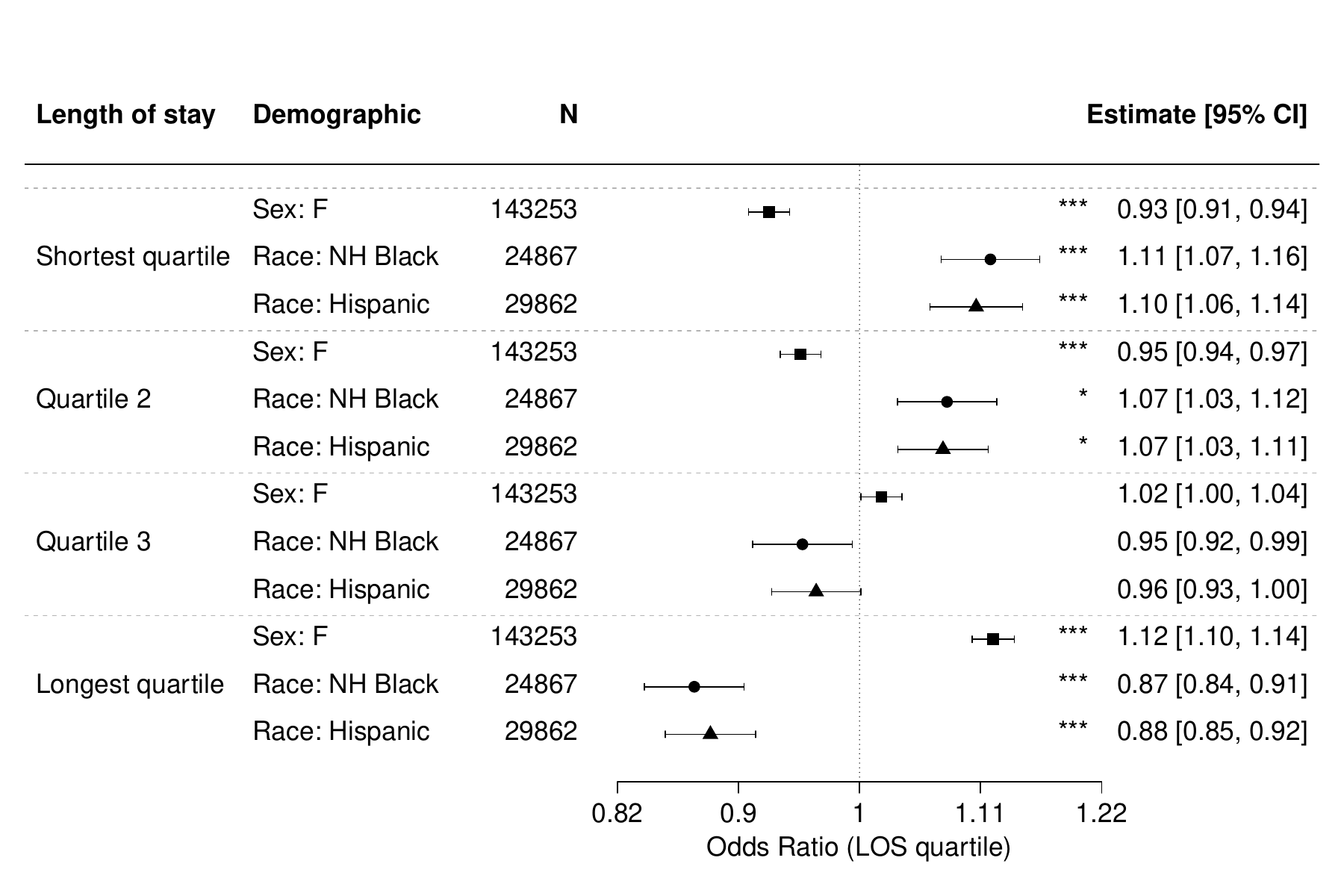}
\includegraphics[trim={11.5cm 0.5cm 0cm 2.0cm},clip, width=0.39\linewidth, height=6cm]{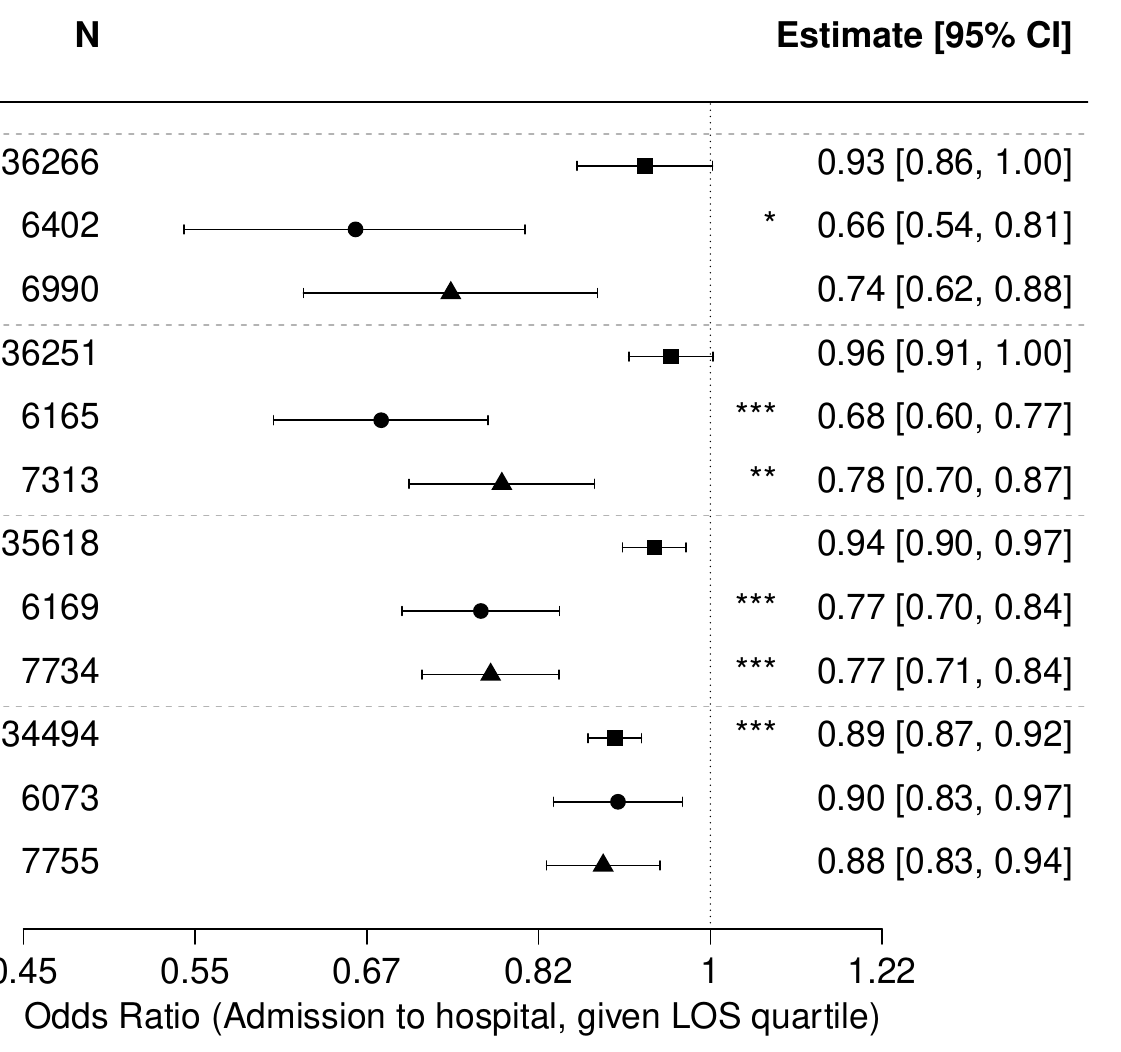}
\caption{Top: Fully adjusted odds ratios describing the likelihood that different demographic groups of patients will be: assigned to each triage group (left); admitted to hospital, given they were assigned to that triage group (right). ESI 5, the least urgent, is excluded due to low sample size. Bottom: Fully adjusted odds ratios describing the likelihood that different demographic groups: will experience different lengths of stay, (left, where lower percentiles indicate a shorter stay), and will be admitted to hospital given each length of stay (right). Odds ratios for female patients are measured relative to male patients; odds ratios for NHB and Hispanic patients are measured relative to NHW patients.}
\label{fig:los-and-ESI-plot}
\end{figure}

Disparities could be observed across scores other than ESI 1 (see Figure \ref{fig:los-and-ESI-plot}), which is reserved for the most severe cases. Female patients were significantly less likely to be assigned a score of 2 than male patients, and more likely to be assigned a score of 3 or 4. Hispanic and NHB patients were less likely than NHW patients to be assigned scores of 2 or 3 and more likely to be assigned a score of 4. We also detected (primarily racial) disparities in the `operation' of ESI scores: Hispanic and NHB patients who had already been assigned an ESI score of 2 or 3 were significantly less likely than NHW patients \textit{at the same triage score} to be admitted to hospital.

Disparities could also be observed in LOS. Female patients tended to have longer stays than male patients, and were less likely to be admitted \textit{only if} they experienced a long stay in the ED; female patients who experienced short stays were admitted at almost the same rates as male patients. Overall, female patients stayed in the ED 26 minutes longer than matched male patients before admission (95\% CI 18-34 min, paired t-test on log-transformed wait times), and 10 minutes longer before discharge (95\% CI 8.5-12 min). When comparing NHB to NHW patients, however, we found the opposite pattern. NHB patients tended to experience\textit{ shorter} stays than NHW patients overall, but racial disparities in admission rates were only significant for patients who experienced shorter stays (see Figure \ref{fig:los-and-ESI-plot}). NHB and Hispanic patients tended to have longer stays before admission than NHW patients (NHB: 40 min., 17-63 min.; H: 26 min., 7.7-44 min.) but were discharged more quickly (NHB: 9.1 min., 4.2-14 min.; H: 8.4 min., 4.0-13 min.)
\subsection{Imaging, Laboratory Tests and Medications} \label{tests-and-meds}

We also examined differences in the rates at which laboratory tests, radiological tests and medications (hereafter collectively referred to as `interventions') are ordered for different groups of patients).

\begin{figure}[htb]
\centering
\includegraphics[trim={0cm 0.75cm 0cm 2.2cm},clip, width=0.6\linewidth, height=6cm]{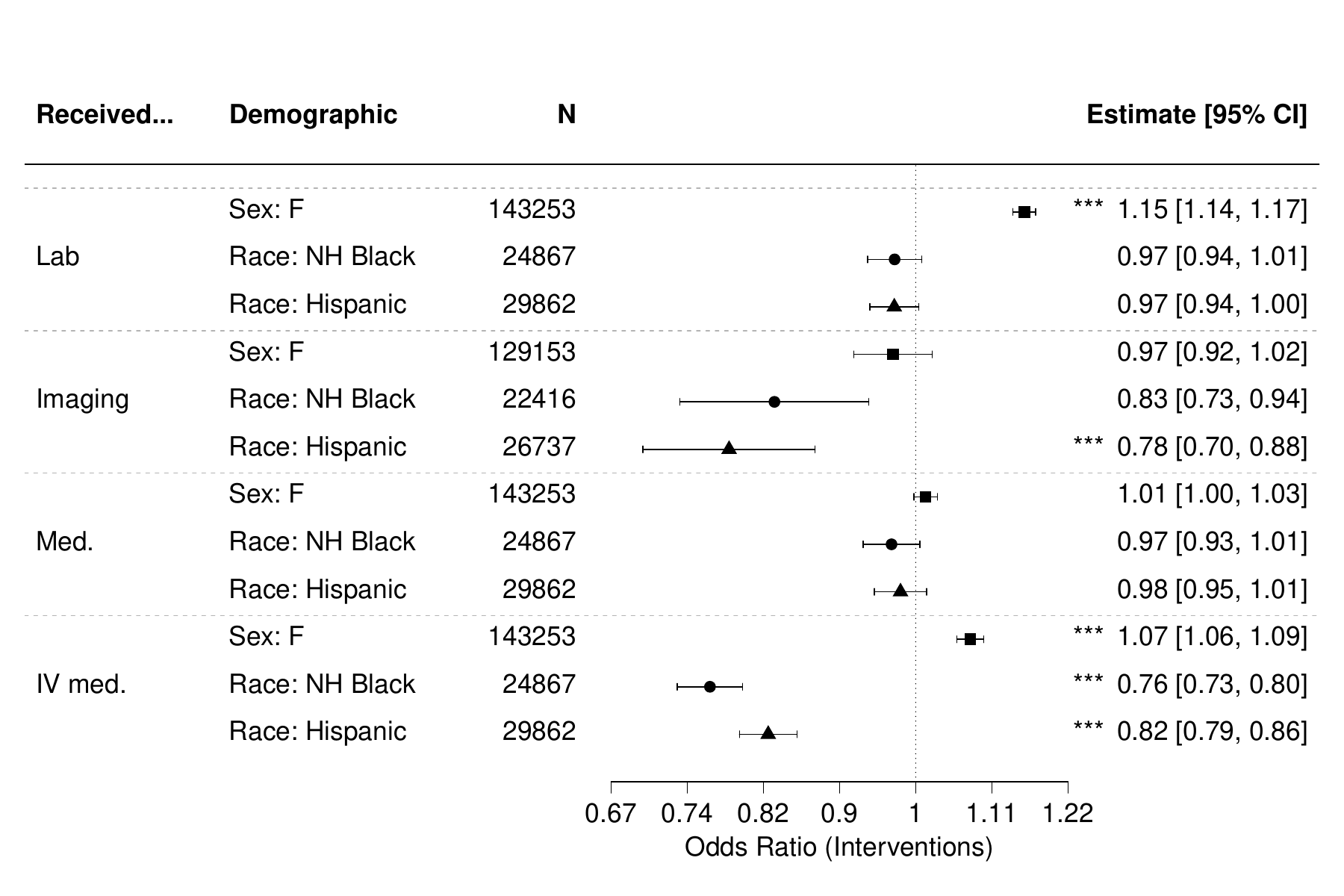}
\includegraphics[trim={11.5cm 0.75cm 0cm 2.2cm},clip, width=0.39\linewidth, height=6cm]{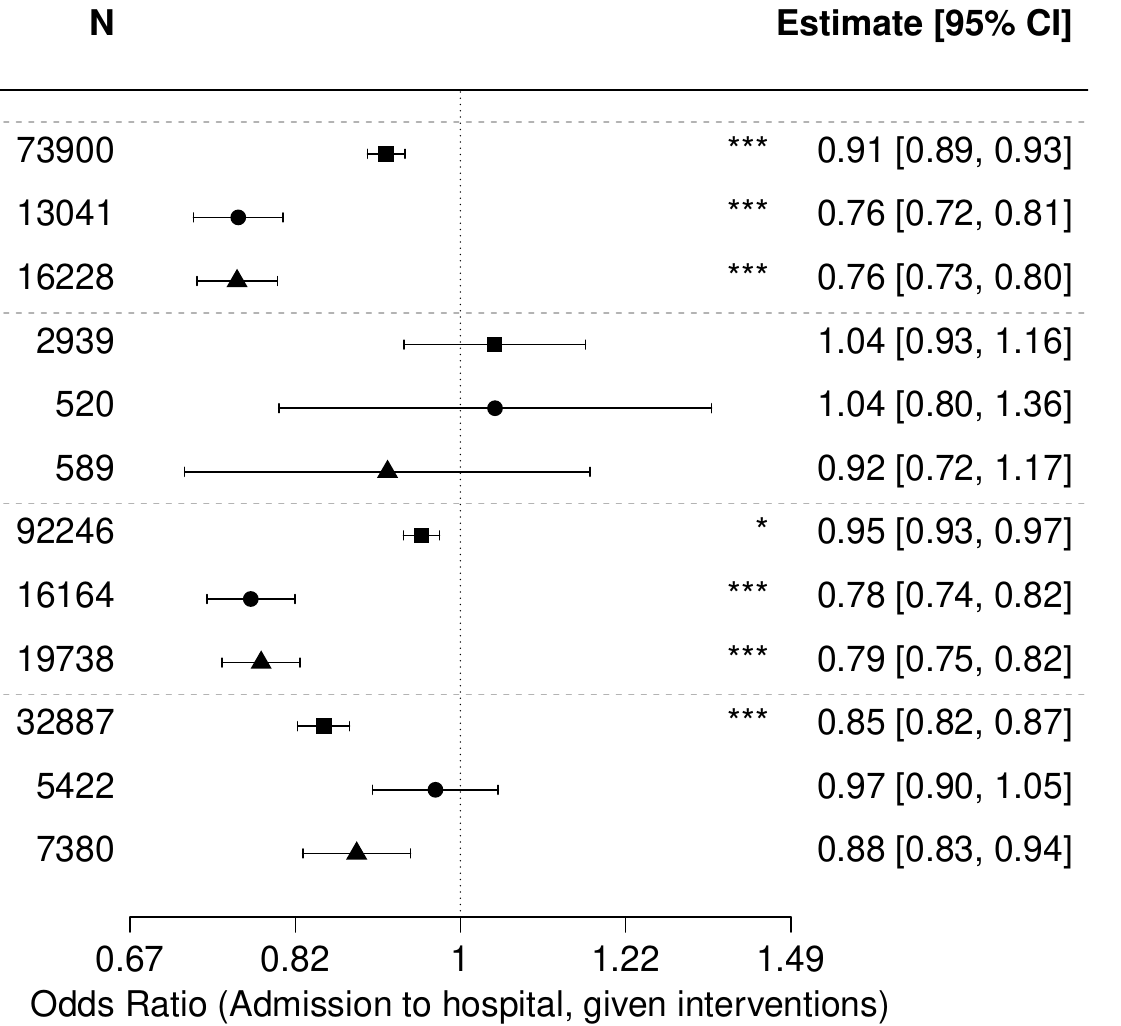}
\caption{Fully adjusted odds ratios describing the likelihood that different demographic groups of patients: received any intervention (left); were admitted to hospital, given they received that intervention (right). Information on laboratory tests were only available if their results came back during the ED visit. `Medications' include IV medications. Imaging refers to radiological tests; we exclude visits after May 2024 for this test, due to difficulties obtaining records. Odds ratios for female patients are measured relative to male patients; odds ratios for NHB and Hispanic patients are measured relative to NHW patients.}
\label{fig:intervention-plot}
\end{figure}

Female patients were significantly more likely than male patients to have laboratory tests ordered for them, and to be treated with intravenous (IV) medications, but were significantly less likely to be admitted after receiving those interventions (see Figure \ref{fig:intervention-plot}). Hispanic patients were significantly less likely to receive radiological imaging, and both NHB and Hispanic patients were significantly less likely to receive IV medications. Whilst we found no significant disparities in the rates at which laboratory tests were ordered, NHB and Hispanic patients who had tests were significantly less likely to be admitted than tested NHW patients. 

\subsection{Vital Signs, Reported Pain, and Crowding} \label{ref-pain-crowding}

\begin{figure}[htb!]
\centering
\includegraphics[width=0.6\linewidth]{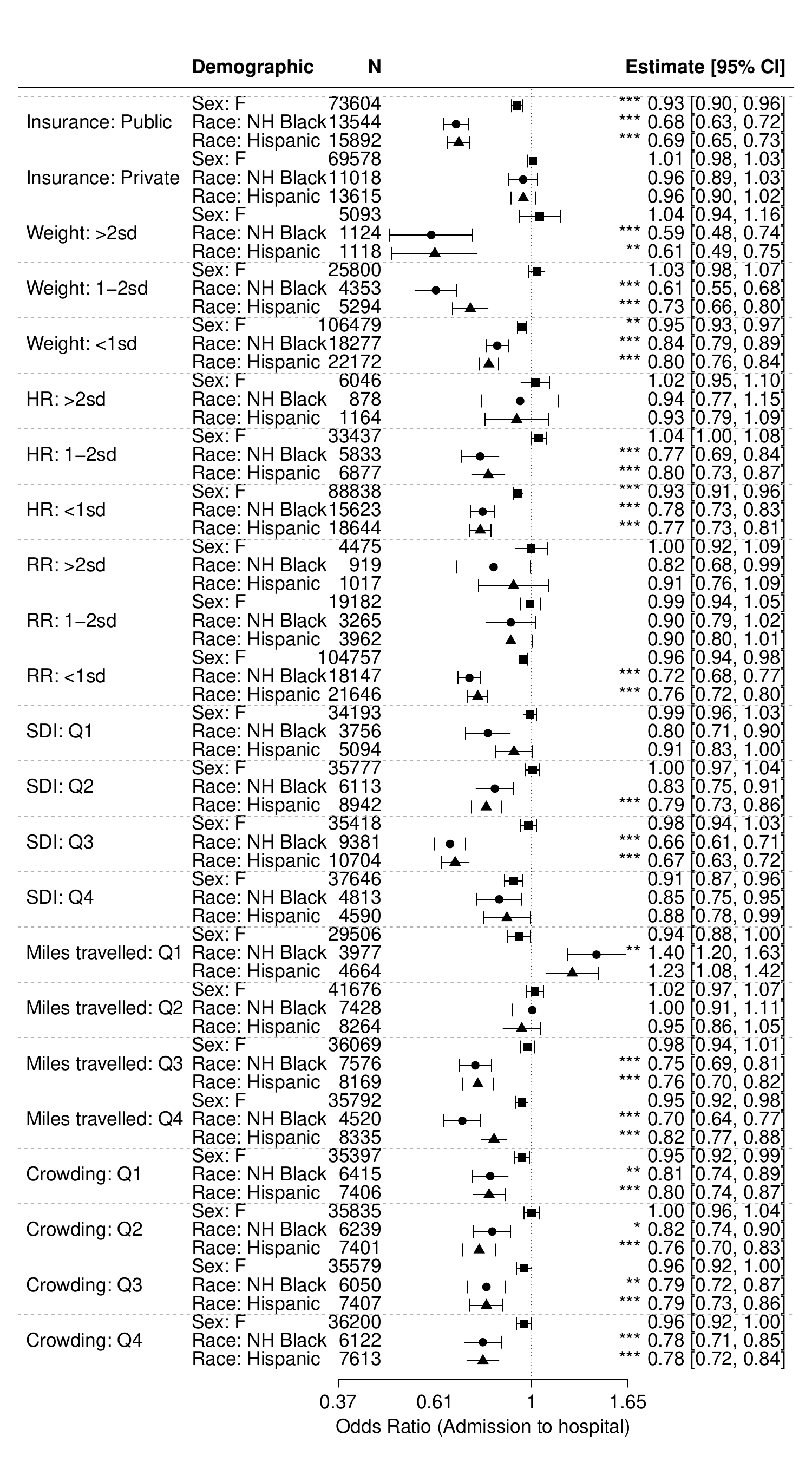}
\caption{Fully adjusted odds ratios describing the likelihood that different demographic groups of patients will be admitted to hospital, given various characteristics. HR = heart rate; O2 = oxygen saturation; SDI = social deprivation index. SDI Q4 is most deprived, SDI Q1 is least. RR, HR and weight are measured in numbers of standard deviations (sd) from the age-group mean, so that lower values indicate `normal' vitals and higher values indicate `abnormal' vitals.}
\label{fig:hr-and-rr-plot}
\end{figure}

We next considered whether visit characteristics outside of the clinician's control, such as insurance type and the vitals measured across a patient's stay, affected disparities in admission. We first divided visits into categories based on these characteristics (e.g. public and private insurance), and then tested those subgroups individually for admission disparities (see Figure \ref{fig:hr-and-rr-plot}). NHB and Hispanic patients with public insurance were less likely than NHW patients with public insurance to be admitted, but no significant disparities existed between patients with private insurance. Racial disparities were significant in the second and third SDI quartiles, but vanished at the fourth (the most deprived quartile). This suggests that `moderate deprivation' tends to increase racial disparities.  NHB patients who lived locally to the hospital were \textit{more} likely than similar NHW patients to be admitted, but NHB patients who lived far away were less likely.

We also observed that disparities decreased as a patient's vital readings (HR, RR and oxygen saturation) became more abnormal for their age group. The exception to this rule was weight: sex-based admission disparities were individually significant only for patients with a `normal' weight, but race-based disparities were \textit{exacerbated} by abnormal weight (see Figure \ref{fig:hr-and-rr-plot}). 

Other subpopulations in which NHB-NHW and Hispanic-NHW disparities were individually significant included: patients aged 5-10 (NHB: OR=0.68, 95\% CI 0.70-0.76, H: 0.79, 0.72-87) and 15+ (NHB: 0.67, 0.61-0.74; H: 0.74, 0.68-0.81); patients brought in by emergency medical services (NHB: 0.73, 0.64-0.83; H: 0.71, 0.63-0.81); walk-in arrivals (NHB: 0.77, 0.73-0.81; H: 0.80, 0.76-0.84); patients with no thirty-day history of admissions (NHB: 0.77, 0.73-0.81; H: 0.76, 0.73-0.80) or visits without admission; (NHB: 0.76, 0.72-0.80; H: 0.77, 0.74-0.80); English-speakers (NHB: 0.75, 0.71-0.78; H: 0.77, 0.74-0.81); patients with previous diagnoses of gastrointestinal conditions (NHB: 0.58, 0.45-0.73; H: 0.53, 0.43-0.65) or nausea (H: 0.71, 0.59-0.83); and patients complaining of fever (NHB: 0.67, 0.56-79), abdominal pain (NHB: 0.48, 0.39-0.60; H: 0.73, 0.62-0.85), vomiting (NHB: 0.46, 0.35-0.61; H: 0.57, 0.45-0.72), or seizures (H: 0.58, 0.45-0.74). Female patients were much less likely than male patients to be admitted with previous diagnoses of nausea (0.80, 0.73-0.86) or when complaining of abdominal pain (0.74, 0.69-0.80), but were much more likely to be admitted for psychiatric complaints (1.37, 1.23-1.53). We found that, after full adjustment, female, NHB and Hispanic patients waited slightly but significantly longer to leave the ED once an admission decision had been taken (female vs male: 12 min., 95\% CI 7.8-17 min.; NHB vs NHW: 20 min., 7.4-33 min.; H vs NHW: 11 min., 1.0-21 min.), suggesting that minoritised patients may be de-prioritised when assigning inpatient beds. 

Finally, to properly understand the effect of language on admission decisions, we matched patients of a variety of languages to English speakers of the same race. We found that NHW Portuguese speakers were significantly less likely to be admitted than NHW English speakers (0.67, 0.55-0.82), as were NHW speakers of `other' languages than the 7 most common specified in Table \ref{tab:demographics} (0.80, 0.70-0.90). However, NHB speakers of Portuguese were significantly \textit{more} likely to be admitted (2.63, 1.52-1.73) than NHB English speakers, suggesting that the effect of language on a patient's experience in the ED is dependent on their race.
\subsection{Impact on Admission Prediction}

\begin{figure}[htb!]
\centering
\includegraphics[width=0.45\textwidth]{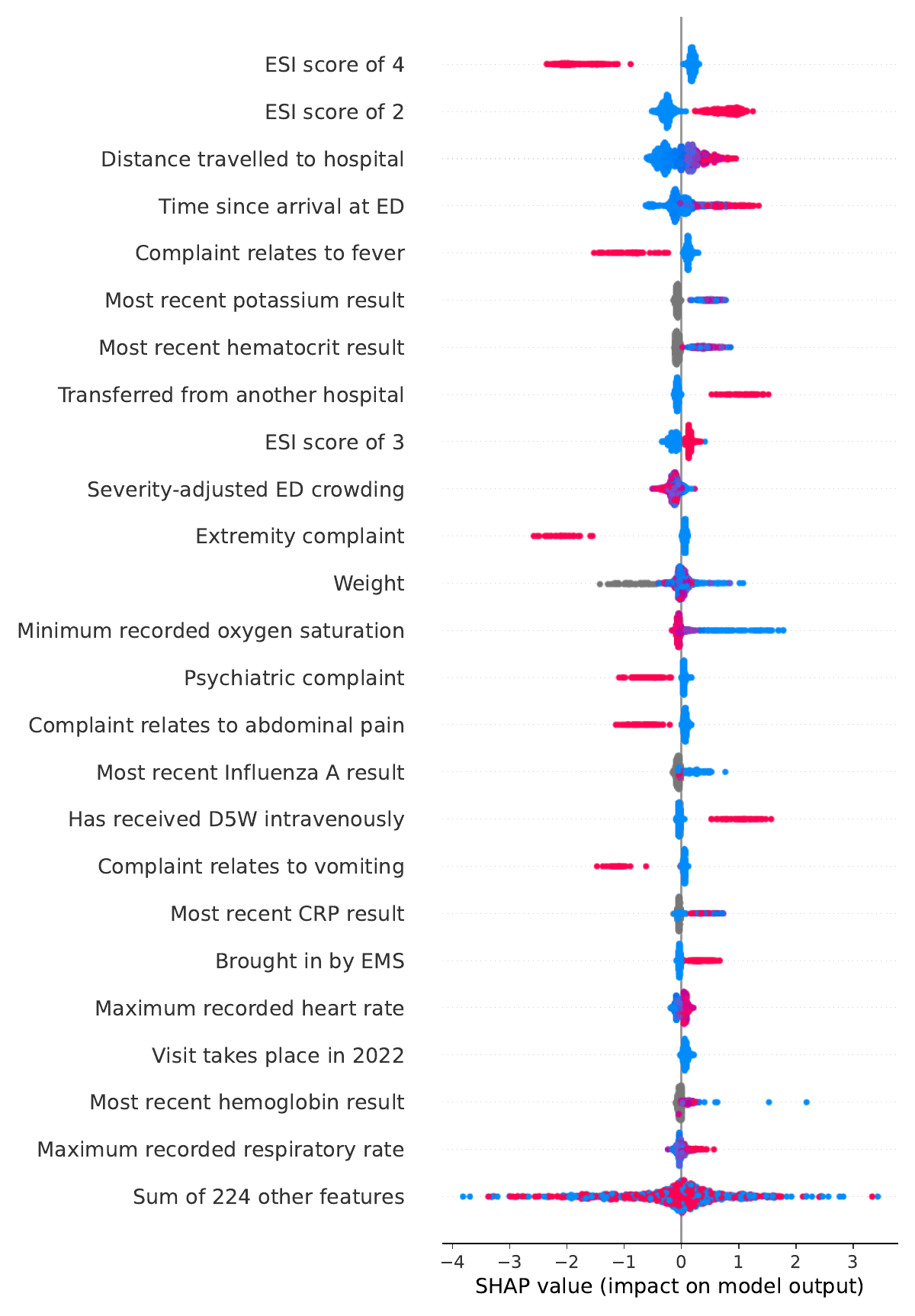}
\raisebox{0.6cm}{\includegraphics[width=0.45\textwidth, height=3.95in]{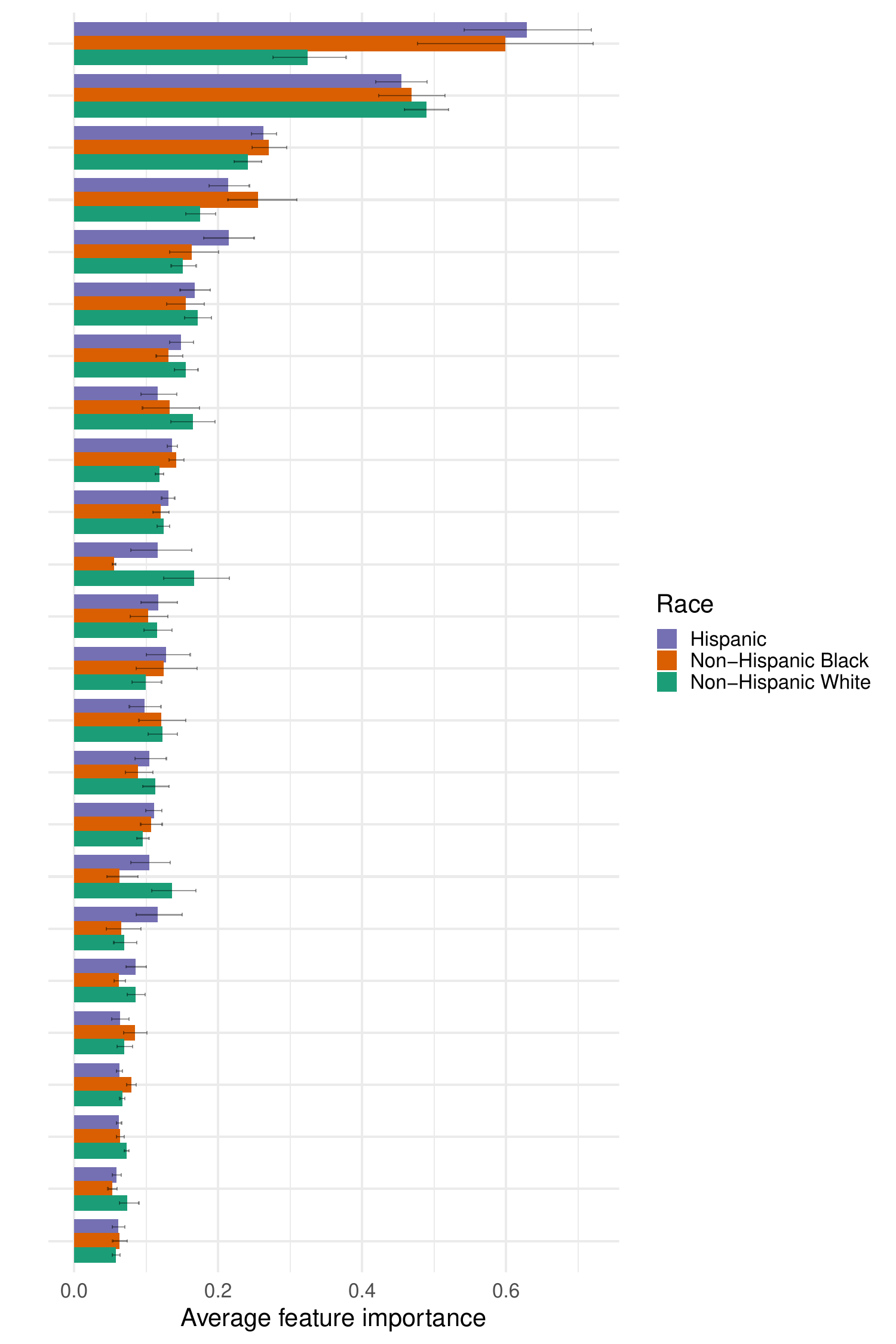}}
\caption{Left:  The top 24 most important features highlighted by our model, which is trained on individual timepoints throughout a patient's visit. Blue dots indicate low feature value, red dots indicate high, and grey indicate that the information is unmeasured (e.g. grey for a laboratory test indicates that, if ordered, its result was unknown at that point in the visit). SHAP values are taken from 1000 random timepoints in the test set. Right: Average feature importance for key features, broken out by patient race. Errorbars indicate bootstrapped 95\% confidence intervals.}
\label{fig:shap-plot}
\end{figure}

To test whether these disparities would affect the workings of clinical decision support tools, we trained a boosted tree model to predict examined disparities in the features treated as important by our model. Using only timepoints taken at the start of each visit, our model had an overall AUROC of 0.921 (bootstrapped 95\% CI 0.918-0.923) and AUPRC of 0.632 (0.620-0.643); at the one-hour mark, an AUROC of 0.928 (0.926-0.930) and AUPRC of 0.700 (0.689-0.708); and at the end of each visit, an AUROC of 0.970 (0.968-0.971) and AUPRC of 0.841 (0.835-0.848). The model relied most heavily on ESI score, the presence of fever, duration of stay, and distance travelled to the hospital (see Figure \ref{fig:shap-plot}).

To clarify how disparities in patient treatment affect the model's ability to learn from the decisions taken during a patient's stay, we broke out SHAP values by race for key features (see Figure \ref{fig:shap-plot}). We found that the model `penalised' NHB and Hispanic patients more heavily for a non-urgent ESI score and `rewarded' NHW patients more an emergent one. Certain other features, such as oxygen saturation and current length of visit, were more important for NHB than NHW patients. These feature importance disparities highlight that models trained on biased data will apply those biases to future cohorts.

\section{Discussion}

Our study observed wider demographic disparities where providers had greater discretion, such as where patients had `normal' vitals, in line with previous studies of radiological testing \cite{Payne2009}. This suggests that the subjective judgement of clinicians about the `look' of a patient may be affecting the treatment of borderline cases and the speed at which admitted patients are prioritised for inpatient beds. Training to address subconscious biases could be effective in reducing racial disparities \cite{Zeidan2019}, especially in the specific clinical subpopulations in which those biases are most acute. These include patients with abnormal weight and from areas of moderate socioeconomic deprivation, as well as speakers of non-English primary languages, whose use of an interpreter may influence a physician's impression of their testimony.

Our results also suggest that female patients may be `overtreated' with IV medications and `overtested' in the ED relative to male patients, leading to longer stays and lower eventual admission rates. These effects, along with differences in the interpretation of vital signs, modified the importance assigned by a machine learning model to clinical decisions when trying to predict admission for minoritised groups. 

Our dataset came from a single site, and our findings may not extrapolate to other hospitals or beyond the pediatric care setting. We did not account for the specific provider or team of providers which treated a particular patient, and did not have a record of the medications a patient was taking at the time of their admission to the ED. Whilst we used the number and severity of patients waiting in the ED as a measure of `crowdedness', we lacked information on NEDOCS scores (which take into account the number of beds available within the hospital and the number of clinicians on staff) and so may not have fully account for the pressures on the ED. Nonetheless, our findings illuminate a number of important patterns in the `mechanisms' of race and sex-based disparities, and should inform both the education of future clinicians and the development of clinical decision support tools.

\section*{Acknowledgments}

Research reported in this publication was supported by the National Library of Medicine of the National Institutes of Health (NIH) Award R01LM014300. 
The content is the responsibility of the authors and does not necessarily represent the official views of NIH.

\bibliographystyle{ws-procs11x85}

\bibliography{ed}
\end{document}